# Strong coupling in the far-infrared between graphene plasmons and the surface optical phonons of silicon dioxide

I. J. Luxmoore[1,a)], C. H. Gan[1], P. Q. Liu[2], F. Valmorra[2], P. Li[1], J. Faist[2], and G. R. Nash[1]

[1]*College of Engineering, Mathematics and Physical Sciences, University of Exeter, Exeter, EX4 4QF, United Kingdom.*

[2]*Institute for Quantum Electronics, ETH Zurich, Wolfgang-Pauli-Strasse 16, CH-8093 Zurich, Switzerland.*

## Abstract

We study plasmonic resonances in electrostatically gated graphene nanoribbons on silicon dioxide substrates. Absorption spectra are measured in the mid-far infrared and reveal multiple peaks, with width-dependent resonant frequencies. We calculate the dielectric function within the random phase approximation and show that the observed spectra can be explained by surface-plasmon-phonon-polariton modes, which arise from coupling of the graphene plasmon to three surface optical phonon modes in the silicon dioxide.

a) E-mail: i.j.luxmoore@exeter.ac.uk

Graphene has been identified as a potential plasmonic material with resonances in the mid-IR to THz region of the electromagnetic spectrum[1–6], which provides a wealth of opportunity for technological exploitation in free space communication, security, bio-sensing and trace gas detection[2]. The ability to electrostatically gate the charge density in graphene to in excess of $1\times10^{13}$cm$^{-2}$ predicates tunable and switchable devices[1,4,7,8] and, coupled with an effective mass which is small compared to that of two dimensional electron gases in conventional semiconductors, results in significantly enhanced light-plasmon coupling and the observation of plasmons at room temperature[1].

To excite graphene plasmons using light it is necessary to overcome the momentum mismatch between the surface-plasmon and free-space photons. This can be achieved using near-field excitation, where graphene plasmons have been launched and imaged and shown to have a high degree of electromagnetic confinement[9,10]. Plasmons can also be observed in optical experiments by patterning the graphene into structures with dimensions from 10s of nm to several microns[1,4,5,11]. Etching graphene into arrays of ribbons allows electromagnetic radiation to excite plasmons with $q \approx (2n + 1)\pi/w$ (where w is the width of the ribbon and $n = 0,1,2 ....$), as the incident electric field induces oscillations in the free charge, resulting in the appearance of absorption features in transmission spectra. In micron scale ribbons, the absorption spectrum is determined purely by the plasmon dispersion, with the resonance frequency, $\omega_{pl} \propto \sqrt{q}$, lying in the terahertz region[1]. When the ribbon width is reduced to the order of several hundred nanometers the absorption spectra become more complicated, with multiple resonances in the mid-IR, arising from coupling between the plasmon and optical phonons in the underlying substrate[5,12].

In this work we investigate the spectrum of graphene plasmons, confined to sub-micron widths, in the mid, but also the far IR. We study the transmission of electrically contacted nanoribbon arrays, which allow control of the graphene Fermi level via an applied back gate voltage. We observe four distinct peaks in the spectra and show using a calculation of the dielectric function within the random phase approximation (RPA) that the behavior can be attributed to the coupled modes, including one previously unobserved, of the graphene plasmon and surface optical phonons of the underlying $SiO_2$ substrate. We also show that the calculations are consistent with a large range of ribbon widths corresponding to the frequency range from the THz to mid-infrared

The devices, pictured in Fig. 1(a) are fabricated from pre-transferred chemical vapor deposited (CVD) monolayer graphene (Graphene Square) on 300nm thick $SiO_2$. The underlying Si substrate is used as a back gate and has a sheet resistance of ~10Ωcm. Electron beam lithography and reactive ion etching are used to pattern the graphene into 300µm × 300µm nanoribbon arrays with widths between 150 and 500nm, where the width is ~40% of the ribbon repeat distance. Cr/Au source and drain contacts are deposited onto the graphene at either end of the nanoribbon array to allow electrical measurement and biasing. The nanoribbons are connected by 500nm wide perpendicular bridges every 10µm to ensure that a small break in an individual nanoribbon does not result in its electrical isolation. Fig. 1(b) shows a scanning electron microscope (SEM) image of a typical graphene nanoribbon device.

Un-patterned graphene devices are also fabricated on the same chip to allow estimation of the mobility and doping of the graphene. Fig. 2(a) shows the field effect characteristic of an un-patterned graphene device, where the resistance of the graphene is measured as a function of the voltage applied to the back gate, $V_G$. This shows behavior typical of monolayer graphene with a peak in resistance at $V_G \approx 88V$, corresponding to the charge neutral point ($V_{CNP}$), where the carrier density is minimized. Fitting the experimental data in the voltage range close to $V_{CNP}$ using a phenomenological model[13] allows us to estimate the mobility, $\mu$ to be $\sim 600 cm^2 V^{-1} s^{-1}$. The large positive value of $V_{CNP}$ indicates that the fabricated samples have a significant intrinsic hole doping, which is also the case for the nanoribbon devices. Fig. 2(b) shows the field effect characteristic for a ribbon array with w=180nm, where $V_{CNP} \approx 80V$. The large intrinsic doping allows wide tuning of the Fermi level, which can be approximated with a simple capacitor model[11] for 300nm thick $SiO_2$ as $|E_F| = 0.031|\sqrt{V_{CNP} - V_G}|$. The resistance is $\sim 5X$ larger than the unpatterned graphene, suggesting that the mobility is reduced, however, it is difficult to confirm whether this is the case as the narrow width and relatively long length between bridges means that many ribbons have breaks at some point along their length, thus contributing to an increased resistance.

Infra-red spectral transmission measurements were performed in air at room temperature using a home-built infrared microscope coupled to a Fourier Transform Infrared Spectrometer (FTIR) with a measurement range of $\sim 250$-$6000 cm^{-1}$. Collimated light from the FTIR is focused onto the sample using a 36X reflective objective. The transmitted light is collected with a second reflective objective and focused onto a pyro-electric detector using an off-axis parabolic mirror. The incident light is linearly polarized using a broadband wire grid polarizer.

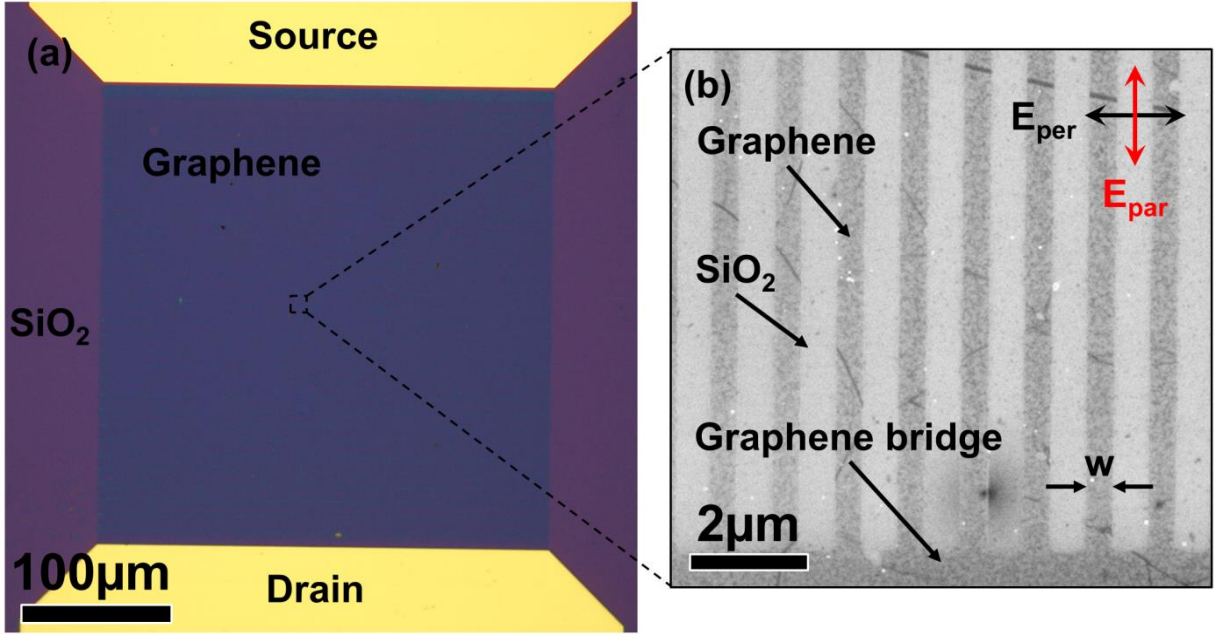

**Fig. 1** (a) Optical and (b) SEM image of graphene nanoribbon array device. In (b) w indicates the graphene ribbon width and $E_{per}$ ($E_{par}$) represents the linear polarization of the incident electromagnetic radiation, where the electric field component is perpendicular (parallel) to the nanoribbon.

In order to investigate absorption resonances arising from plasmons in the graphene nanoribbons, transmission spectra are measured for a given Fermi level. In the following, we present results for $E_F$=-0.37eV. The extinction is defined as $1 - T/T_{CNP}$, where $T$ is the transmission spectrum at $E_F$=-0.37eV and $T_{CNP}$ is the transmission spectrum at the charge neutrality point, $V_G = V_{CNP}$. Such an extinction spectrum is shown in Fig. 2(c) for a nanoribbon with *w*=180nm. When the incident light is polarized parallel to the nanoribbons ($E_{par}$ in Fig.1 (b)) the absorption is not strongly affected by the graphene Fermi level. The only feature in the spectrum is a small increase in the absorption at low frequencies due to the Drude absorption[1]. When the incident light is polarized perpendicular to the nanoribbons ($E_{per}$ in Fig.1 (b)), the absorption is very different, with four sharp absorption resonances revealed, labelled $P_1$-$P_4$ in Fig. 2(c). These absorption resonances can be attributed to the interaction of graphene plasmons with surface optical (SO) phonons[14] in the underlying $SiO_2$ substrate. Their coupling via the long range Fröhlich interaction[15] results in the excitation of surface-plasmon-phonon-polariton (SP3) modes[5,11,12]. In previous work, peaks $P_2$-$P_4$ were observed[5,12] due to SO phonon modes at 806 and 1168cm$^{-1}$. Here, we report the observation of strong coupling between the SO phonon mode at 485cm$^{-1}$ and the graphene plasmon mode, which results in the additional resonance, $P_1$.

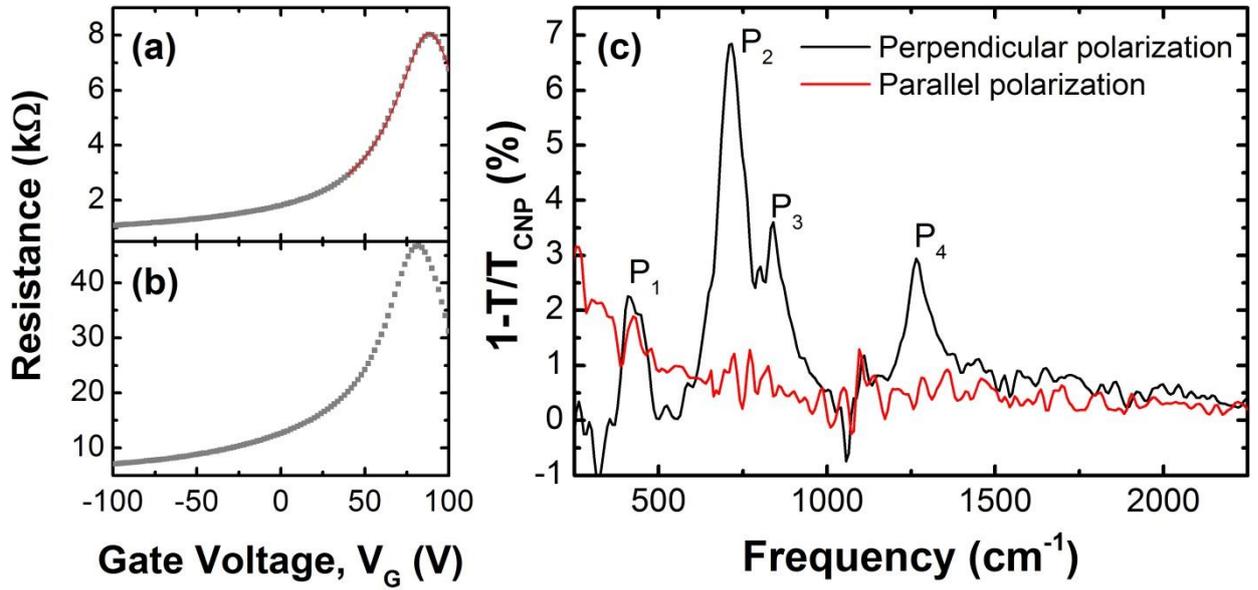

**Fig. 2** Field effect characteristic of (a) unpatterned graphene device and (b) graphene nanoribbon array with ribbon width of 180nm. The red line in (a) is a fit to the experimental data following ref. 13. (c) Extinction spectra of 180nm wide graphene nanoribbon array on $SiO_2$ for incident light polarized parallel and perpendicular to the ribbons. When the light is polarized perpendicular to the nanoribbon, four peaks labelled $P_1$-$P_4$, are clearly identified.

To confirm the origin of $P_1$, we investigate the absorption spectrum of a range of different nanoribbon widths between 150 and 500nm. If $P_1$ does originate from the coupling of the plasmon resonance and a substrate SO phonon, then an anti-crossing with $P_2$ should be observed as the plasmon wavevector, $q = \pi/(w - w_0)$ is tuned by the width of the ribbon, where $w_0$ is an experimentally determined parameter which accounts for damage at the edge of the ribbons caused by the reactive ion etch[5,16,17]. Fig. 3(a) plots the extinction spectra for several devices with different ribbon widths between 180 and 450nm. As the ribbon width is increased, all four peaks shift to lower frequency but at different rates, with $P_1$ and $P_2$ shifting the most. The spectra are fit with a Fano model[5,18] and the extracted peak positions are plotted in Fig. 3(b) as a function of $q$. We take $w_0 = 0$ as the edge damage has been shown to vary considerably[5,16,17] and we do not determine it experimentally for our devices. The data suggests that there is an anti-crossing of $P_1$ and $P_2$ at ~485cm$^{-1}$ and to confirm that this arises from the plasmon-phonon coupling we calculate the dispersion and loss function of the SP3 modes for graphene on a SiO$_2$ substrate.

The SO phonon frequencies are first obtained by solving the dispersion relation $1 + \epsilon_p(\omega) = 0$, where

$$\epsilon_p(\omega) = \epsilon_0 + \sum_{n=1}^{N} \frac{f_n \omega_{TO,n}^2}{\omega_{TO,n}^2 - \omega^2}$$

(1)

is the ionic dielectric function of an insulator exhibiting $N$ SO phonon modes[19], $\omega$ is the frequency, $\omega_{TO,n}$ is the frequency of the $n^{th}$ transverse optical (TO) phonon mode and it is assumed that $\omega_{TO,n+1} > \omega_{TO,n}$. In Eq. (1), $f_n = \epsilon^{(n-1)} - \epsilon^{(n)}$ is the oscillator strength of the $n^{th}$ mode, such that $\sum_{n=1}^{N} f_n = \epsilon_s - \epsilon_\infty$, with $\epsilon_s = \epsilon^{(0)}$ being the static dielectric constant and $\epsilon_\infty = \epsilon^{(N)}$ the high frequency dielectric constant. It is taken that $\epsilon_s$=3.9 and $\epsilon_\infty$=2.4 for the SiO$_2$ substrate. The values of $\epsilon^{(n)}$ that appear in the oscillator strength $f_n$ were approximated as[19] $\epsilon^{(n)} \approx \epsilon^{(n-1)} \frac{\omega_{TO,n}^2}{\omega_{LO,n}^2}$, and the dispersion calculated from Eq. (1) for three SO modes (*N*=3). In the calculations, the transverse optical (TO) and longitudinal optical (LO) phonon frequencies are taken to be $\omega_{TO}$=[448, 791.7, 1128.1] cm$^{-1}$ and $\omega_{LO}$=[498.6, 811.5, 1317] cm$^{-1}$, following values obtained from the literature[19,20], and the SO frequencies ($\omega_{SO} \approx \omega_{LO,n}\sqrt{\frac{\epsilon^{(n)}(\epsilon^{(n-1)}+1)}{\epsilon^{(n-1)}(\epsilon^{(n)}+1)}}$) were found to be [484.8, 805.9, 1229.0] cm$^{-1}$. The SO mode energies are plotted as dotted lines in Fig. 3(b) and agree well with previous measurements[5,19] and our experimental data. The corresponding ratios of the oscillator coupling strength $f_2/f_1$=1/5 and $f_3/f_1$=4/5 are also in close agreement to our experimental data and previously reported values[5,19].

To obtain the dispersion of the hybrid SP3 modes, we consider the coupling between a two dimensional electron gas (the graphene plasmons) and dispersionless SO phonons at zero temperature[21]. Zero-temperature calculations are expected to yield quantitatively accurate results for our measurements taken at room temperature (T~300K, $K_BT$~26meV), since $E_F$>>$K_BT$ and the lowest SO phonon mode energy considered is $\hbar\omega_{SO}^{(min)} \gtrsim$ 60meV. Within the random phase approximation, the total dielectric function is[5,21]

$$\epsilon_{tot}(q,\omega) = 1 - v_F \Pi - \frac{1}{1+\beta}$$

(2)

where $v_F = \frac{e^2}{2q\epsilon_\infty\epsilon_0}$ is the 2D Fourier transform of the Coulomb potential, $\Pi$ is the polarizability of the uncoupled graphene sheet as calculated in ref. 22, and

$$\beta = \left[ \epsilon_0 e^{-2qd} \sum_{n=1}^{N} \frac{\alpha_n \omega_{SO,n}^2}{\omega^2 - \omega_{SO,n}^2} \right]^{-1}$$

(3)

where the weighted coupling coefficient $\alpha_n = \frac{f_n}{(\epsilon^{(n-1)}+1)(\epsilon^{(n)}+1)}$ of each mode is proportional to the Fröhlich coupling strength[21] and $d$ is the distance between the graphene and the substrate, taken to be 3.5Å in the calculations. The loss function is obtained by taking the imaginary part of the inverse dielectric function, ie. $L = -\Im \frac{1}{\epsilon_{tot}}$ (refs. 15,21). The RPA loss function is plotted, along with the uncoupled graphene dispersion and the peak positions from the experimental data, in Fig. 3(b). We see good agreement between the experimental data and the calculations, confirming that the measured spectra arise from the SP3 modes originating from the coupling of graphene plasmon with three SO phonon modes in the $SiO_2$ substrate. We additionally plot in Fig. 3(b) the peak positions from micro-ribbon arrays with widths of 1 and 2μm, fabricated using the same procedure, to show the good agreement between calculations and experiment over a wide frequency range from the THz to mid-infrared.

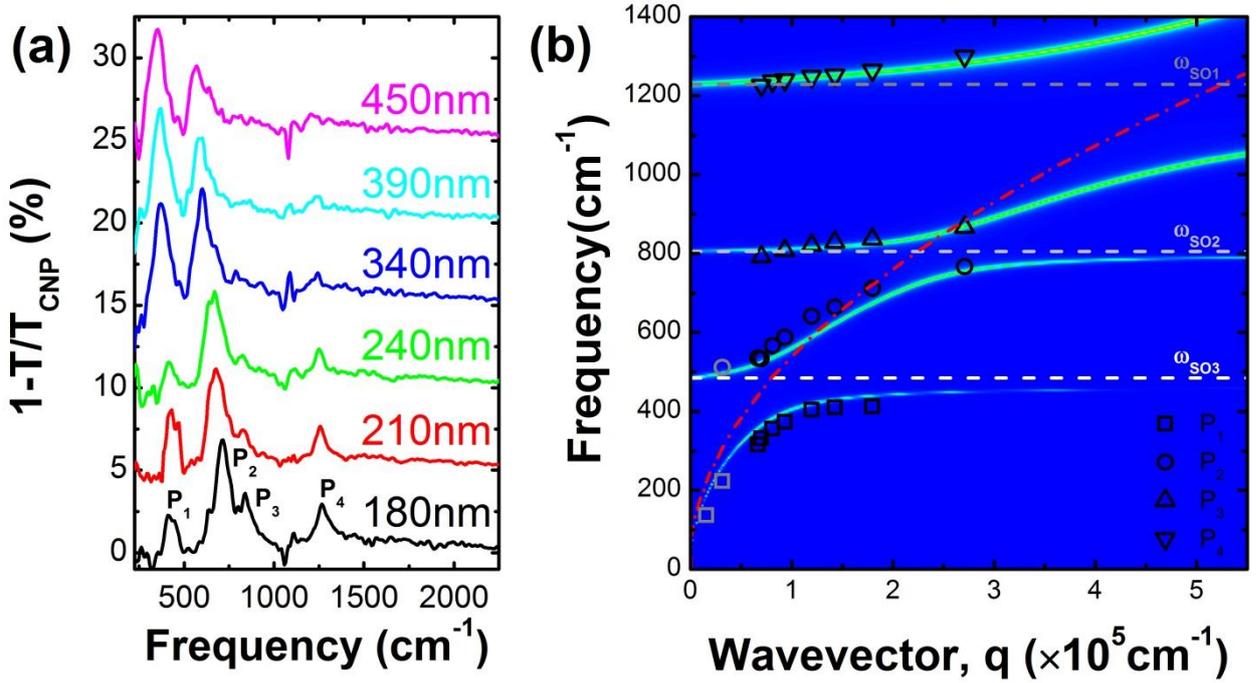

**Figure 3.** (a) Extinction spectra measured for graphene nanoribbons with a range of widths. (b) Calculated loss plot with extracted peak frequencies overlaid. The red dashed line shows the calculated dispersion of the uncoupled graphene plasmon. The grey, light grey and white dashed lines show the calculated frequency of the three surface optical phonons, $\omega_{SO1}$=1229cm$^{-1}$, $\omega_{SO2}$=806cm$^{-1}$, $\omega_{SO3}$=485cm$^{-1}$. The grey symbols are peak positions extracted from micro-ribbon arrays and show that the calculations are consistent with a large range of ribbon widths which span a frequency range from the THz to mid-infrared.

In the local limit ($\omega \gg \tau^{-1}$ and $\omega \gtrsim v_F q$, with a phenomenological relaxation time for electrons), one may adopt the Drude model to describe the conductivity of monolayer graphene as $\sigma(\omega) = \frac{ie^2 E_F}{\pi \hbar^2 (\omega + i\tau^{-1})}$, which leads to a simplified form of the polarizability[2,6]

$$\Pi \approx \Pi^{Drude} \approx \frac{2q^2 \epsilon_\infty E_F}{\pi^2 \hbar^2 (\omega^2 + \omega \tau^{-1})}$$

(4)

In general, the dynamical conductivity of graphene has two contributions, intra-band (Drude) and inter-band (see, e.g., equations (4) - (7) in ref. 23). Including the inter-band contribution is essential when $\hbar\omega > 2E_F$. In our case $\hbar\omega < 2E_F$, therefore the inter-band conductivity is not taken into account. Based on calculations of the dispersion of the SP3 modes with $\Pi^{Drude}$ in Eq. 2, the Drude approximation was found to be highly accurate for $\omega > 350$cm$^{-1}$ and nanoribbon widths $w \geq 100$nm.

The lifetime of the SP3 modes can be extracted from the spectral linewidth and we measure values of <100fs for all four resonances, comparable to previous measurements[5]. For P$_4$, the lifetime shows a monotonic increase from ~40fs to ~80fs as the nanoribbon width is reduced from 480nm to 180nm, indicating that the effect of edge scattering is insignificant in this regime. The increase in lifetime in this range can be explained by the convergence of the resonant frequency of the SP3 mode and the SO phonon frequency at 485cm$^{-1}$, whereby the SP3 mode becomes mode phonon like[5]. The energy is far below the graphene optical phonon energy (~0.2eV), yet despite the lack of edge effects the lifetime is still short compared to the SO phonon lifetime of ~1ps. This suggests that the lifetime is dominated by damping mechanisms intrinsic to the graphene, such as impurity[24] and defect scattering[25], which is consistent with the low mobility, typical of CVD graphene and high doping shown in Fig. 2(b).

In conclusion, we have experimentally observed plasmonic resonances in graphene nanoribbons. The broad measurement range allows the simultaneous observation of four absorption peaks corresponding to the surface-plasmon-phonon-polariton modes arising from the Fröhlich coupling of graphene plasmons and the three SO phonon modes with frequencies of ~485, 806 and 1230cm$^{-1}$. Coupling to the SO phonon at 485cm$^{-1}$ pushes graphene plasmonics into the technologically relevant far infrared regime and with improvements in material quality, graphene plasmonics can play an important role in future nano-photonic devices.

**Acknowledgements**

This research was supported by the UK Engineering and Physical Sciences Research Council and the European Union under the FET-open grant GOSFEL. [13]

**References**

[1] L. Ju, B. Geng, J. Horng, C. Girit, M. Martin, Z. Hao, H. A. Bechtel, X. Liang, A. Zettl, Y. R. Shen and F. Wang, Nat. Nanotechnol. **6**, 630 (2011).

[2] T. Low and P. Avouris, ACS Nano **8**, 1086 (2014).

[3] A. Grigorenko, M. Polini, and K. Novoselov, Nat. Photonics (2012).

[4] H. Yan, X. Li, B. Chandra, G. Tulevski, Y. Wu, M. Freitag, W. Zhu, P. Avouris, and F. Xia, Nat. Nanotechnol. **7**, 330 (2012).

[5] H. Yan, T. Low, W. Zhu, Y. Wu, M. Freitag, X. Li, F. Guinea, P. Avouris and F. Xia, Nat. Photon. **7**, 394 (2013).

[6] C. H. Gan, Appl. Phys. Lett. **101**, 111609 (2012).

[7] Y. V. Bludov, M.I. Vasilevskiy, and N.M.R. Peres, EPL Europhys. Lett. **92**, 68001 (2010).

[8] H.-S. Chu and C. H. Gan, Appl. Phys. Lett. **102**, 231107 (2013).

[9] Z. Fei, A. S. Rodin, G. O. Andreev, W. Bao, A. S. McLeod, M. Wagner, L. M. Zhang, Z. Zhao, M. Thiemens, G. Dominguez, M. M. Fogler, A. H. Castro Neto, C. N. Lau, F. Keilmann and D. N. Basov, Nature **487**, 82 (2012).

[10] J. Chen, M. Badioli, P. Alonso-González, S. Thongrattanasiri, F. Huth, J. Osmond, M. Spasenović, A. Centeno, A. Pesquera, P. Godignon, A. Zurutuza Elorza, N. Camara, F. J. García de Abajo, R. Hillenbrand and F. H. L. Koppens, Nature **487**, 77 (2012).

[11] V. Brar, M. S. Jang, M. Sherrott, J. J. Lopez, and H. A. Atwater, Nano Lett. **13**, 2541 (2013).

[12] M. Freitag, T. Low, W. Zhu, H. Yan, F. Xia, and P. Avouris, Nat. Commun. **4**, 1951 (2013).

[13]V. E. Dorgan, M. -H. Bae, and E. Pop, Appl. Phys. Lett. **97**, 082112 (2010).

[14]R. Fuchs and K. Kliewer, Phys. Rev. **140**, 2076 (1965).

[15]S. Q. Wang and G. D. Mahan, Phys. Rev. B **6**, 4517 (1972).

[16]C. Berger, Z. Song, X. Li, X. Wu, N. Brown, C. Naud, D. Mayou, T. Li, J. Hass, A. N. Marchenkov, E. H. Conrad, P. N. First, and W. A de Heer, Science **312**, 1191 (2006).

[17]M. Han, B. Özyilmaz, Y. Zhang, and P. Kim, Phys. Rev. Lett. **98**, 206805 (2007).

[18]Z. Li, C. H. Lui, E. Cappelluti, L. Benfatto, K. F. Mak, G. L. Carr, J. Shan, and T. F. Heinz, Phys. Rev. Lett. **108**, 156801 (2012).

[19]M. V. Fischetti, D. A. Neumayer, and E. A. Cartier, J. Appl. Phys. **90**, 4587 (2001).

[20]P. Umari, A. Pasquarello, and A. Dal Corso, Phys. Rev. B **63**, 094305 (2001).

[21]E. H. Hwang, R. Sensarma, and S. Das Sarma, Phys. Rev. B **82**, 195406 (2010).

[22]E. H. Hwang and S. Das Sarma, Phys. Rev. B **75**, 205418 (2007).

[23]S. A. Mikhailov and K. Ziegler, Phys. Rev. Lett. 99 , 016803 (2007).

[24]T. Langer, J. Baringhaus, H Pfnür, H.W. Schumacher and C. Tegenkamp, New J. Phys.**12** 033017 (2010).

[25]A. Principi, G. Vignale, M. Carrega and M. Polini, Phys. Rev. B. **88** 121405(R) (2013).